\documentclass[%
 reprint,
superscriptaddress,
 amsmath,amssymb,
 aps,
]{revtex4-1}

\usepackage{amsmath}
\usepackage{graphicx}% Include figure files
\usepackage{dcolumn}% Align table columns on decimal point
\usepackage{bm}% bold math
\usepackage{hyperref}% add hypertext capabilities
\begin{document}

\preprint{APS/123-QED}

\title{Destroying the Event Horizon of Cold Dark Matter-Black Hole System
}%

\author{Liping Meng}
% \altaffiliation{College of Physics, Guizhou University, Guiyang 550025, China}%Lines break automatically or can be forced with \\
\author{Zhaoyi Xu}
 %\email{ }
 \author{Meirong Tang}
 \email{Electronic address: tangmr@gzu.edu.cn(Corresponding author)
}
\affiliation{%
College of Physics, Guizhou University,\\
 Guiyang 550025, China
}%





\date{\today}% It is always \today, today,

\begin{abstract}
Since the weak cosmic censorship conjecture was proposed, research on this conjecture has been ongoing. This paper explores the conjecture in black holes that are closer to those existing in the real universe (i.e., rotating black holes enveloped by dark matter). In this paper, we obtained a first-order corrected analytical solution for the black hole event horizon through an approximate solution. The validity of the first-order corrected analytical solution will be provided in the appendix. We conduct our study by introducing a test particle and a scalar field into the black hole. Our conclusions show that, in extremal case, both a test particle and a scalar field can disrupt the event horizon of the Kerr-like black hole; in near-extremal case, both a test particle and a scalar field can disrupt the event horizon of the Kerr-like black hole. When cold dark matter is not considered, the conclusion is consistent with previous research.
\begin{description}
\item[Keywords]
Cold dark matter, Weak cosmic censorship conjecture, The event horizon, Analytical solution
\end{description}
\end{abstract}

\maketitle

%\tableofcontents
\section{Introduction}
According to Newtonian mechanics, when calculating the rotational speeds $\nu ^{2}=\frac{GM}{r}$ of stars in galaxies, it is evident from the formula that the rotational speed of stars on the outer edges of galaxies decreases as their distance from the galaxy's mass center increases. However, it has been found that the rotational speeds of stars in NGC 3198, NGC 6503, and the Milky Way remain nearly constant at distances several times that of the galactic center\cite{Begeman:1989kf,Bottema:1997ya}. Therefore, researchers have hypothesized the existence of dark matter (DM)\cite{Zwicky:1933gu,Rubin:1980zd}, and currently, DM has been confirmed by various indirect evidences\cite{Gammaldi:2017mio,Blandford:1991xc,Bramante:2023djs}. As a result, extensive research has been conducted on dark matter, leading to the proposal of various models, such as the Cold Dark Matter (CDM) model\cite{Navarro:1995iw,Navarro:1996gj}, Warm Dark Matter model\cite{Colin:2000dn,Bode:2000gq}, Bose-Einstein Condensate (BEC) Dark Matter model\cite{Hu:2000ke,Press:1989id,Turner:1983he}, and Self-Interacting Dark Matter model\cite{Spergel:1999mh,Zeng:2023fnj}. In the mainstream cosmological model, Cold Dark Matter (CDM) makes a significant contribution to the total density of the universe, accounting for 26.8$\%$ of it\cite{Planck:2019nip}. Research in this area will promote breakthroughs in astrophysics, particle physics, and cosmology.
In the "standard" Lambda Cold Dark Matter ($\Lambda$CDM) cosmological model, the density distribution of dark matter on the scale of galaxies is described by a universal form proposed by Navarro, Frenk, and White, known as the NFW profile\cite{Navarro:1995iw,Navarro:1996gj}\cite{Navarro:1994hi}. Its mathematical representation is as follows
\begin{equation}\label{1}
\rho _{NFW}=\frac{\rho_{s} }{\frac{r}{R_{s}} (1+\frac{r}{R_{s} })^{2} },
\end{equation}
here, the critical density $\rho_{s}$ and the scale radius $R_{s}$ are referenced. The NFW profile is universally applicable to cosmological models dominated by cold dark matter across different mass ranges. However, on smaller scales, the Einasto profile would be a better choice\cite{Navarro:2003ew,Navarro:2008kc}.

The connection between black holes and dark matter is an intriguing subject of current research. In theory, the dark matter halos that surround black holes can influence them. Studies have indicated that dark matter in the vicinity of black holes can form a spiking phenomenon\cite{Gondolo:1999ef,Sadeghian:2013laa}, and that the presence of dark matter facilitates the formation of black holes\cite{Lora-Clavijo:2014kha,Zelnikov:2003du,Munyaneza:2004ai,Volonteri:2005fj}. Relevant studies have also been documented in the literature\cite{LeDelliou:2009twa,LeDelliou:2009gqa}. Considering the weak cosmic censorship conjecture, which posits the non-existence of naked singularities and suggests that singularities are enclosed by the event horizons of black holes\cite{Penrose:1969pc}, exploring the existence and potential issues of dark matter and black holes is a topic worthy of in-depth research. The immersion of a black hole in a dark matter halo could potentially disrupt its event horizon. If the weak cosmic censorship conjecture is violated, the issue of the existence of black holes becomes particularly noteworthy.

The Weak Cosmic Censorship Conjecture was proposed by Penrose\cite{Penrose:1969pc}, and its core idea is that singularities must be hidden behind the event horizon, preventing distant observers from directly observing them. This ensures the causal structure and predictive power of general relativity. Although the conjecture has not yet been universally proven with a rigorous mathematical formulation within the framework of general relativity, reference\cite{Wald:1997wa} provides a modern mathematical expression of the conjecture. At present, this conjecture has been widely studied through various forms of numerical simulations, such as the numerical evolution of matter field collapse\cite{East:2019bwu,Song:2020onc,Goswami:2007na}, extensive non-linear numerical simulations of black holes\cite{Li:2020smq,Shaymatov:2022ako,Choptuik:2003qd}, and the numerical evolution of black holes in multiple dimensions\cite{Ahmed:2022dpu,Nie:2021rhz,Sperhake:2009jz,Andrade:2020dgc,Zhang:2020txy}.
Wald showed through Gedanken experiments\cite{Wald:1974hkz,Sorce:2017dst} that, in certain situations, the weak cosmic censorship conjecture may be applicable to the studied black holes. Based on this method, other researchers further demonstrated that the conjecture is applicable to other types of black holes\cite{Chen:2019nhv,Zhang:2020txy,Gwak:2019rcz,Yang:2020czk,Zhao:2023vxq}. However, studies in references\cite{Jacobson:2009kt,Rocha:2014jma,Saa:2011wq,Hod:2019wcw} showed that under certain specific conditions, the weak cosmic censorship conjecture may be violated, and thus its applicability still deserves further exploration. It is worth noting that this paper considers test particles (or test scalar fields) that have no impact on the curved black hole spacetime. However, if the nonlinear influence of particles (or scalar fields) on the curved spacetime itself is considered, the weak cosmic censorship conjecture may still hold in this type of gedanken experiments. In this context, the work of Hod \cite{Hod:2002pm,Hod:2008zza,Hod:2013oxa} has provided important insights in this field. Through precise analysis, he has shown that after considering the effect of particle self - energy, the integrity of the black hole horizon can still be guaranteed. These studies cleverly utilized the Hawking’s area theorem and  self-energy corrections, demonstrating that even under extremal conditions, the weak cosmic censorship conjecture still holds. These research findings not only provide strong support for the verification of this conjecture but also offer profound implications for further understanding black hole physics.

General relativity predicts the existence of black holes\cite{LIGOScientific:2016aoc}, which are composed of event horizons and gravitational singularities within these horizons\cite{EventHorizonTelescope:2019dse,EventHorizonTelescope:2019ggy}. The validity of the weak cosmic censorship conjecture is crucial for the theoretical existence of black holes and is a prerequisite for their reasonable definition. Xu and others have calculated a Schwarzschild-like and Kerr-like black hole metric solutions immersed in a dark matter halo under asymptotic conditions and a constant state equation\cite{Xu:2020jpv}. Based on this work, this paper analyzes the existence of event horizons in the Kerr-like black hole by injecting a test particle and a scalar field into it, thereby exploring the applicability of the weak cosmic censorship conjecture in the black hole. For ease of calculation, this paper adopts the natural unit system.

The paper is structured as follows: In Sect.2, the properties of the studied Kerr-like black holes are introduced, including the event horizon structure equations and the rotational angular velocity of the black hole. In Sects.3 and 4, a test particle and a classical scalar field are respectively introduced into the Kerr-like black holes to study the weak cosmic censorship conjecture in these black holes. In Sect.5, we discuss the issues encountered in the research and provide a summary.

\section{Kerr-Like Black Holes Immersed in a Dark Matter Halo}
Considering the isotropy of the pressure of the dark matter halo, as in $P_{r} =P_{\theta }=P_{\phi }$, the approximate metric solution for a black hole immersed in a dark matter halo can be obtained in the Boyer–Lindquist coordinates as described in $(t,r,\theta ,\phi )$\cite{Xu:2020jpv},
\begin{equation}\label{2}
\begin{split}
ds^2=&-\frac{r^2g(r)+a^2{cos}^2{\theta}}{\mathrm{\Sigma}^2}dt^2+\frac{\mathrm{\Sigma}^2}{r^2g(r)+a^2}dr^2\\&-\frac{2a{sin}^2{\theta}(k(r)-r^2g(r))}{\mathrm{\Sigma}^2}d\phi dt+\mathrm{\Sigma}^2d\theta^2\\&+\mathrm{\Sigma}^2{sin}^2{\theta}[1+a^2{sin}^2{\theta}\frac{2k(r)-r^2g(r)+a^2{cos}^2{\theta}}{\mathrm{\Sigma}^4}]d\phi ^2.
\end{split}
\end{equation}
In the metric
\begin{equation}\label{3}
g(r)=1-\frac{B(r)}{r},
\end{equation}
\begin{equation}\label{4}
B(r)=2M+8\pi\rho_sR_s^3[ln{(}1+\frac{r}{R_s})+\frac{R_s}{r+R_s}-1],
\end{equation}
wherein
\begin{equation}\label{5}
\mathrm{\Sigma}^2=k(r)+a^2{cos}^2{\theta},
\end{equation}
\begin{equation}\label{6}
k(r)\approx r^2,
\end{equation}
\begin{equation}\label{7}
\mathrm{\Delta}=r^2g(r)+a^2.
\end{equation}
$M$ represents the mass of the black hole, and $a$ is the spin parameter (with $J=Ma$, $J$ denoting the angular momentum of the black hole). If the dark matter halo does not exist, the above metric of the black hole would reduce to that of a Kerr black hole.

Under this metric, the structural equation for the black hole's event horizon is the following equation
\begin{equation}\label{8}
r^2-2Mr-8\pi\rho_sR_s^3r[ln{(}1+\frac{r}{R_s})+\frac{R_s}{r+R_s}-1]+a^2=0.
\end{equation}
This is a transcendental equation, so an analytical solution cannot be obtained. Here, we will use an approximate method to solve it. For convenience, we set
\begin{equation}\label{9}
\sigma=8\pi\rho_sR_s^3.
\end{equation}
Since dark matter has a minimal effect on the black hole event horizon, for $\sigma \left [ ln(1+\frac{r}{R_{s}})+\frac{R_{s}}{r+R_{s}}-1   \right ] \ll 1$ near the event horizon, we can represent the event horizon of such Kerr black holes as a modified form of the standard Kerr black hole event horizon.

In metric \eqref{2}, when the dark matter halo is absent, that is, when $\sigma=0$, the Kerr-like metric degenerates into a Kerr black hole, with its event horizon and Cauchy horizon given by
\begin{equation}\label{10}
r_{0+} =M+\sqrt{M^{2}-a^{2} },
\end{equation}

\begin{equation}\label{11}
r_{0-} =M-\sqrt{M^{2}-a^{2} }.
\end{equation}
Here, the ‘+’ and ‘-’ denote the event horizon and the Cauchy horizon of the Kerr black hole, respectively. The parameters $r_{0+}$ and $r_{0-}$ are considered as the zeroth-order solution (unmodified solution) for the Kerr-like black hole. Next, we introduce the correction terms for the dark matter halo to obtain
\begin{equation}\label{12}
\bigtriangleup _{kerr}-\sigma r\left [ ln(1+\frac{r}{R_{s}})+\frac{R_{s}}{r+R_{s}}-1   \right ] =0,
\end{equation}
since the correction terms due to the dark matter halo are small, the above expression can be approximated as
\begin{equation}\label{13}
\bigtriangleup _{kerr}-\sigma r\left [ ln(1+\frac{r}{R_{s}})+\frac{R_{s}}{r+R_{s}}-1   \right ] \approx (r-r_{1+})(r-r_{1-}).
\end{equation}
Clearly, $r_{1+}$ and $r_{1-}$ represent the first-order correction solutions for the event horizon of the black hole system with a dark matter halo, and are given by
\begin{equation}\label{14}
r_{1+}=M+\sqrt{M^2-a^2+\sigma r_{0+}\left [ ln(1+\frac{r_{0+}}{R_{s}})+\frac{R_{s}}{r_{0+}+R_{s}} -1  \right ] } ,
\end{equation}

\begin{equation}\label{15}
r_{1-}=M-\sqrt{M^2-a^2+\sigma r_{0-}\left [ ln(1+\frac{r_{0-}}{R_{s}})+\frac{R_{s}}{r_{0-}+R_{s}} -1  \right ] }. 
\end{equation}

By substituting equations \eqref{10} and \eqref{11} into the above expression, we obtain the specific first-order analytical solution for the Kerr-like black hole event horizon. The expression is given by
\begin{widetext}
\begin{equation}\label{16}
r_{1+}=M+\sqrt{ M^2-a^2+\sigma (M+\sqrt{M^2-a^2}) \left [ ln(1+\frac{M+\sqrt{M^2-a^2}}{R_{s}})+\frac{R_{s}}{M+\sqrt{M^2-a^2}+R_{s}} -1  \right ] } ,
\end{equation}
\begin{equation}\label{17}
r_{1-}=M-\sqrt{M^2-a^2+\sigma (M-\sqrt{M^2-a^2}) \left [ ln(1+\frac{M-\sqrt{M^2-a^2}}{R_{s}})+\frac{R_{s}}{M-\sqrt{M^2-a^2}+R_{s}} -1  \right ] } .
\end{equation}
\end{widetext}

In this paper, we first use the first-order correction analytical solution \eqref{16} to explore whether the weak cosmic censorship conjecture is violated. This is because when using a test particle or a scalar field, the variations of our the test particle and scalar field are first-order small quantities. Since the conditions discussed and the expressions for the event horizon here are of the same order, it is reasonable to discuss them in this way initially. To assess the validity of using the first-order analytical solution, we will provide the error range in the appendix (see Appendix ). Of course, if the final analysis necessitates considering second-order effects, we can incorporate the second-order correction solution accordingly. For the second-order correction solution, repeating the above iterative process yields a more accurate event horizon radius as
\begin{equation}\label{18}
r_{2+}=M+\sqrt{M^2-a^2+\sigma r_{1+}\left [ ln(1+\frac{r_{1+}}{R_{s}})+\frac{R_{s}}{r_{1+}+R_{s}} -1  \right ] },
\end{equation}
\begin{equation}\label{19}
r_{2-}=M-\sqrt{M^2-a^2+\sigma r_{1-}\left [ ln(1+\frac{r_{1-}}{R_{s}})+\frac{R_{s}}{r_{1-}+R_{s}} -1  \right ] }.
\end{equation}
Substituting $r_{1+}$ and $r_{1-}$ into the above expression yields the second-order correction analytical solution for the event horizon of a black hole system with a dark matter halo.

Analyzing the equation \eqref{16}, we obtain that the condition for the existence of a black hole event horizon is $a^2 \le M^2$; if the black hole event horizon does not exist, then $a^2>M^2$, that is $J>M^2$.

From the metric of the black hole, we can determine that the black hole is a rotating black hole, hence its angular velocity of rotation is
\begin{equation}\label{20}
\mathrm{\Omega}_h=\frac{a}{k(r_h)+a^2}.
\end{equation}

\section{Overspinning the Black Hole by Injecting a Test Particle}
In this section, we explore the possibility of disrupting the event horizon of a Kerr-like black hole by injecting a test particle into it. Based on the structural equation for the black hole's event horizon, we can derive that the condition for the non-existence of the black hole's event horizon is as follows
\begin{equation}\label{21}
J>M^2.
\end{equation}

Therefore, to obtain the internal structure of the event horizon of the Kerr-like black hole, this paper introduces a test particle with a large angular momentum and a classical scalar field into the black hole. This allows the black hole to absorb energy from the test particle or scalar field, forming a composite system that disrupts the event horizon of the black hole. The formed composite system satisfies condition 
\begin{equation}\label{22}
J\prime>M^{\prime2}.
\end{equation}

In the spacetime of a black hole immersed in a dark matter halo, we use geodesics to describe the trajectory equation of a test particle with a mass of $m$ and the equation can be expressed as
\begin{equation}\label{23}
\frac{d^2x^\mu}{d\tau^2}+\mathrm{\Gamma}_{\alpha\beta}^\mu\frac{dx^\alpha}{d\tau}\frac{dx^\beta}{d\tau}=0.
\end{equation}
The representation form of the Lagrangian for testing particle is as follows
\begin{equation}\label{24}
L=\frac{1}{2}mg_{\mu\nu}\dot{x}^{\mu } \dot{x}^{\nu}.
\end{equation}

When testing particle is incident from infinity on the equatorial plane, there is no velocity component in the $\theta$ direction, so the equation is given by
\begin{equation}\label{25}
P_\theta=\frac{\partial L}{\partial\dot{\theta } }=mg_{22}\dot{\theta} =0,
\end{equation}
by analyzing the motion of the test particle, we obtain the energy $\delta E$ and angular momentum $\delta J$ of the test particle
\begin{equation}\label{26}
\delta E=-P_t=-\frac{\partial L}{\partial\dot{t} }=-mg_{0\nu}\dot{x}^{\nu},
\end{equation}
\begin{equation}\label{27}
 \delta J=P_\phi =\frac{\partial L}{\partial\dot{\phi } }=mg_{3\nu}\dot{x}^{\nu}.
\end{equation}

When a Kerr-like black hole captures a test particle, the energy and angular momentum of the black hole will become
\begin{equation}\label{28}
M\rightarrow M\prime=M+\delta E,
\end{equation}
\begin{equation}\label{29}
J\rightarrow J\prime=J+\delta J.
\end{equation}

According to our study, this essentially means determining the condition under which a test particle can enter the Kerr-like black hole. Then, we refer to our calculated conditions that allow for the disruption of the black hole's event horizon. By combining these two sets of conditions, if the test particle satisfy both, then the internal conditions of the event horizon can be exposed to distant observers.

For a test particle with a mass of $m$, its four-dimensional velocity is described as $\left | \vec{\nu} \right | <c$, and its direction can be described with a unit vector $\vec{\mu}$, that is, it is represented as
\begin{equation}\label{30}
U^\mu U_\mu=g_{\mu\nu}\frac{dx^\mu}{d\tau}\frac{dx^\nu }{d\tau}=\frac{1}{m^2}g^{\mu\nu}P_\mu P_\nu=-1,
\end{equation}
combining \eqref{26}, \eqref{27} and \eqref{30}, we obtain
\begin{equation}\label{31}
\delta E=\frac{g^{03}}{g^{00}}\delta J\pm\frac{1}{g^{00}}\sqrt{(g^{03})^2\delta J^2-g^{00}(g^{33}\delta J^2+g^{11}P_r^2+m^2)}.
\end{equation}

Since the spacetime outside the event horizon is regular, therefore there is
\begin{equation}\label{32}
\frac{dt}{d\tau}>0,
\end{equation}
expanding \eqref{26} and \eqref{27}, we get
\begin{equation}\label{33}
\dot{t}=\frac{dt}{d\tau}=-\frac{g_{33}\delta E+g_{03}\delta J}{g_{00}g_{33}-g_{03}^2}.
\end{equation}
When considering equation \eqref{32}, we have
\begin{equation}\label{34}
\delta E>-\frac{g_{03}}{g_{33}}\delta J,
\end{equation}
so, in \eqref{31}, the value that satisfy the condition of the above equation is
\begin{equation}\label{35}
\delta E=\frac{g^{03}}{g^{00}}\delta J-\frac{1}{g^{00}}\sqrt{(g^{03})^2\delta J^2-g^{00}(g^{33}\delta J^2+g^{11}P_r^2+m^2)}.
\end{equation}

Clearly, for a test particle to enter a black hole, its angular momentum should not be excessively large. This is because greater angular momentum results in a stronger centrifugal repulsion force, which keeps the test particle away from the black hole. Therefore, the upper limit of angular momentum can be given by equation \eqref{34}.
\begin{equation}\label{36}
\delta J<\frac{k(r_h)+a^2}{a}\delta E=\frac{1}{\mathrm{\Omega}_h}\delta E=\delta J_{max}.
\end{equation}

Additionally, for a test particle entering a black hole to disrupt the event horizon of the black hole, there is another condition that must be satisfied, the angular momentum $\delta J$ should have a lower limit. The lower limit is given by equation \eqref{22}. Combining equations \eqref{22}, \eqref{28}, and \eqref{29}, we obtain
\begin{equation}\label{37}
\delta J > \delta J_{min} = (M^{2} - J) + 2M\delta E + \delta E^2,
\end{equation}
therefore, if the test particle satisfies both equations \eqref{36} and \eqref{37}, the event horizon of the Kerr-like black hole will be disrupted by the test particle.

For a Kerr-like black hole, if its initial state is extremal, then $a=M$. In this case, the event horizon is represented as
\begin{equation}\label{38}
r_{1+}=M+\sqrt{\sigma M\left [ ln(1+\frac{M}{R_{s}})+\frac{R_{s}}{M+R_{s}} -1  \right ] } ,
\end{equation}
the angular velocity of the black hole simplifies to
\begin{widetext}
\begin{equation}\label{39}
\Omega _{h}=\frac{1}{2M+\sigma \left [ ln(1+\frac{M}{R_{s}})+\frac{R_{s}}{M+R_{s}}-1   \right ] +2\sqrt{\sigma M \left[ ln(1+\frac{M}{R_{s}})+\frac{R_{s}}{M+R_{s}}-1 \right ]}}.
\end{equation}
\end{widetext}

When energy $\delta E$ is taken to the first-order case, the test particle must simultaneously satisfy the following two conditions for the event horizon to be disrupted at this time.
\begin{widetext}
\begin{equation}\label{40}
\delta J<\delta J_{max}=\frac{1}{\mathrm{\Omega}_h}\delta E=\left \{ 2M+\sigma \left [ ln(1+\frac{M}{R_{s}})+\frac{R_{s}}{M+R_{s}}-1   \right ] +2\sqrt{\sigma M \left[ ln(1+\frac{M}{R_{s}})+\frac{R_{s}}{M+R_{s}}-1 \right ]} \right \}\delta E ,
\end{equation}
\begin{equation}\label{41}
\delta J>\delta J_{min}=2M\delta E.
\end{equation}
\end{widetext}
Combining the above two equations yields
\begin{widetext}
\begin{equation}\label{42}
\delta J_{max}-\delta J_{min}=\left \{\sigma \left [ ln(1+\frac{M}{R_{s}})+\frac{R_{s}}{M+R_{s}}-1   \right ] +2\sqrt{\sigma M \left[ ln(1+\frac{M}{R_{s}})+\frac{R_{s}}{M+R_{s}}-1 \right ]} \right \}\delta E ,
\end{equation}
\end{widetext}
for the sake of convenience in analysis, let
\begin{equation}\label{43}
k=ln(1+\frac{M}{R_{s}})+\frac{R_{s}}{M+R_{s}}-1.
\end{equation}

\begin{figure}[h]
\includegraphics[width=0.5\textwidth]{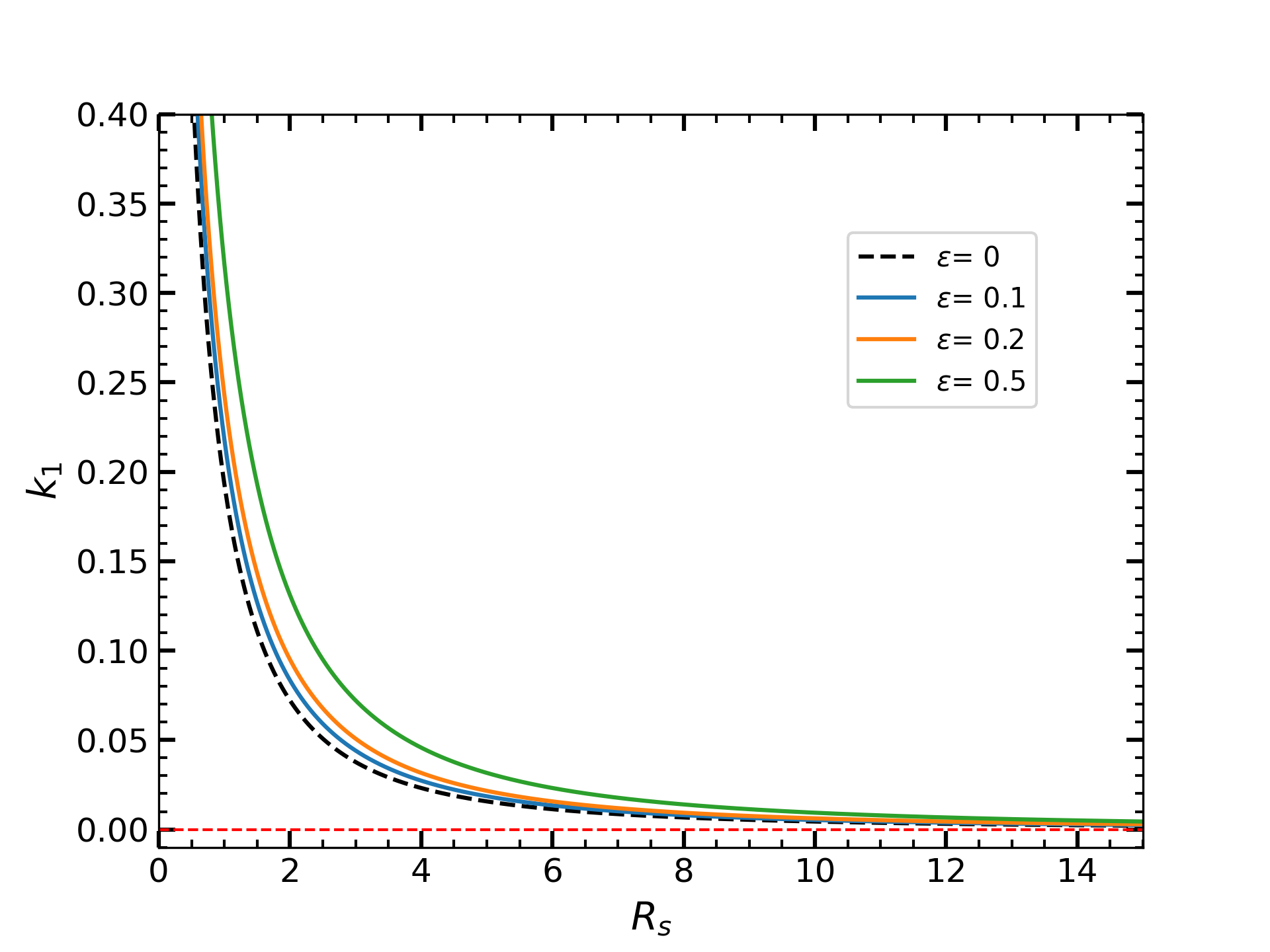}
\caption{The relationship between the critical radius $R_{s}$ and $k_{1}$. When the dimensionless parameter $\varepsilon =0$, the corresponding $k_{1}=k$, which is represented by the black dashed line in the figure. The other curves represent the values of $k_{1}$ under different dimensionless values. Evidently, the values of $k_{1}$ are all greater than zero, and as $R_{s}$ increases, $k_{1}$ gradually approaches zero, i.e., it approaches the red dashed line in the figure. Here, $M=1$.}
\label{a}
\end{figure}

Here, $k>0$, (as shown in Figure \ref{a}) is equivalent to $\sigma k\ge 0$, thus deriving $\delta J_{max}-\delta J_{min} \ge 0$. we find that in extremal case, the test particle can disrupt the event horizon of a Kerr-like black hole. When parameter $\sigma =0$ equals $\delta J_{max}-\delta J_{min}=0$, the Kerr-like black hole degenerates into a Kerr black hole, at which point the test particle cannot disrupt the event horizon of an extremal Kerr black hole.If we take the second-order situation into account, our conclusion remains unchanged. This is because the introduction of the second order only corrects the accuracy range and does not affect our conclusion.

If the initial state of the Kerr-like black hole is near-extremal, and the energy $\delta E$ of the test particle is taken to the first-order term, then we can obtain the condition under which the test particle enters the black hole and is able to disrupt the event horizon of the black hole as follows
\begin{equation}\label{44}
\delta J<\delta J_{max}=\frac{k(r_h)+a^2}{a}\delta E=\frac{1}{\Omega _{h}} \delta E,
\end{equation}
\begin{equation}\label{45}
\delta J>\delta J_{min}=(M^2-J)+ 2M\delta E.
\end{equation}

Define a dimensionless small quantity $\varepsilon$ to describe the degree of deviation from the extremal condition in the near-extremal case, namely
\begin{equation}\label{46}
\frac{a^2}{M^2}=1-\varepsilon^2,
\end{equation}
the analysis of the above equation shows that when the dimensionless small quantity $\varepsilon$ approaches zero, it describes a near-extremal situation; when the dimensionless small quantity $\varepsilon$ equals zero, it describes an extremal situation.

In the first-order approximation, combining equations \eqref{44} and \eqref{45}, the condition that can destroy the black hole event horizon is equivalent to
\begin{equation}\label{47}
(\frac{1}{\mathrm{\Omega}_h}-2M)\delta E-(M^2-J)>0,
\end{equation}
calculating $M^2-J$ yields
\begin{equation}\label{48}
M^2-J=\frac{M^2 \varepsilon ^2}{2} +O(\varepsilon ^4).
\end{equation}
Evidently, $M^2-J$ is a second-order infinitesimal, thus \eqref{47} can be equivalent to
\begin{equation}\label{49}
\frac{1}{\mathrm{\Omega}_h}-2M>0.
\end{equation}
According to equations \eqref{16} and \eqref{46}, we obtain
\begin{widetext}
\begin{equation}\label{50}
\frac{1}{\Omega _{h}}-2M=k_{1}\sigma +2\sqrt{h_{1}}M+k_{1}\varepsilon +\left ( \frac{k_{1}\sigma }{2}+2M+\sqrt{h_{1}}M  \right )\varepsilon ^{2}+\frac{k_{1}}{2}\varepsilon ^{3} +O(\varepsilon ^{4}),
\end{equation}
\end{widetext}
where
\begin{equation}\label{51}
k_{1}=ln\left ( 1+\frac{M+\varepsilon }{R_{s}}  \right )+\frac{R_{s}}{M+\varepsilon +R_{s}}-1,
\end{equation}
\begin{equation}\label{52}
h_{1}=\varepsilon ^2+\frac{\sigma (1+\varepsilon )\left [ ln\left ( 1+\frac{M+\varepsilon }{R_{s}}  \right )+\frac{R_{s}}{M+\varepsilon +R_{s}}-1  \right ] }{M}.
\end{equation}

Analyzing the above expression, we find that $k_{1}>0$ (as shown in Figure \ref{a}). Therefore, we can conclude that equation \eqref{49} holds, so in near-extremal conditions, a test particle can indeed destroy the event horizon of a Kerr-like black hole. This means that the weak cosmic censorship conjecture is violated in this black hole spacetime. 
It is worth noting that, our above conclusion does not consider the backreaction effects of spacetime, this is because the main contribution of the spacetime background or self-interaction effects come from electromagnetic effects, which are higher-order effects\cite{Sorce:2017dst} and our a test particle or a scalar field do not carry any charge (neutral test bodies). Therefore, the impact on our conclusions is not very significant. Of course, in the future, we will fully consider spacetime effects through numerical simulations to obtain an accurate range of test particles that can destroy the event horizon.

In addition, we note that when the parameter is $\sigma =0$, the Kerr-like black hole degenerates into a Kerr black hole. By analyzing the expression \eqref{50}, it can be concluded that the event horizon of a near-extremal Kerr black hole may be destroyed by a test particle ($\frac{1}{\Omega _{h} }-2M>0 $). This is consistent with the conclusion in reference \cite{Hod:2002pm}, that is, in the case of ignoring the self-energy effect, the test particle may cause the near-extremal Kerr black hole to over-spin, thus destroying its event horizon. However, the reference also points out that the self-energy effect plays an important role in protecting the event horizon under certain near-extremal conditions. The self-energy effect of particles can prevent the black hole from over-spinning under certain conditions by modifying its interaction with the black hole. Therefore, in future work, further exploring the impact of the self-energy effect on the stability of black holes will contribute to a more comprehensive understanding of the applicability of the weak cosmic censorship conjecture.

\section{a Classical Scalar Field Entering a Black Hole}
In this section, we discuss the possibility of the event horizon of a Kerr-like black hole being disrupted in extremal and near-extremal cases, by injecting a scalar field with large angular momentum into the Kerr-like black hole.
\subsection{ A Massive Classical Scalar Field}
In this subsection, we study the mass scattering between a classical scalar field and a Kerr-like black hole immersed in a dark matter halo, where the mass of the scalar field $\mathrm{\Psi}$ is $\mu$. The equation of the scalar field $\mathrm{\Psi}$ is given by the Klein-Gordon equation
\begin{equation}\label{53}
\frac{1}{\sqrt{-g}}\partial_\mu(\sqrt{-g}g^{\mu\nu}\partial_\nu\mathrm{\Psi})-\mu^2\mathrm{\Psi}=0.
\end{equation}
Substituting the metric \eqref{2} of a Kerr-like black hole into the above equation yields 
\begin{equation}\label{54}
\begin{split}
&-\frac{(k(r)+a^2)^2-a^2\mathrm{\Delta}{sin}^2{\theta}}{\mathrm{\Delta}\mathrm{\Sigma}^2}\frac{\partial^2\mathrm{\Psi}}{\partial t^2}-\frac{2a\eta}{\mathrm{\Delta}\mathrm{\Sigma}^2}\frac{\partial^2\mathrm{\Psi}}{\partial t\partial\phi }\\&+\frac{1}{\mathrm{\Sigma}^2}\frac{\partial}{\partial r}(\mathrm{\Delta}\frac{\partial\Psi}{\partial r})+\frac{1}{\mathrm{\Sigma}^2sin{\theta}}\frac{\partial}{\partial\theta}(sin{\theta}\frac{\partial\Psi}{\partial\theta})\\&+\frac{\mathrm{\Delta}-a^2{sin}^2{\theta}}{\mathrm{\Delta}\mathrm{\Sigma}^2{sin}^2{\theta}}\frac{\partial^2\mathrm{\Psi}}{\partial\phi^2}-\mu^2\mathrm{\Psi}=0.
\end{split}
\end{equation}
To facilitate obtaining the solution of the above equation, the method of separation of variables can be used\cite{Brahma:2020eos}, as follows
\begin{equation}\label{55}
\mathrm{\Psi}(t,r,\theta,\phi)=e^{-i\omega t}R(r)S_{lm}(\theta)e^{im\phi},
\end{equation}
where $S_{lm}(\theta)$ represents the spherical function, and $l,m$ is the constant for separation of variables.

Substituting the above equation into equation \eqref{54} and simplifying and separating variables, we obtain the angular part of the scalar field's equation of motion is 
\begin{equation}\label{56}
\begin{split}
&\frac{1}{sin{\theta}}\frac{d}{d\theta}[sin{\theta}\frac{dS_{lm}(\theta)}{d\theta}]\\&-[a^2\omega^2sin^2\theta +\frac{m^2}{sin^2\theta}+\mu ^2a^2cos^2\theta -\lambda_{lm}]S_{lm}=0,
\end{split}
\end{equation}
the radial part of the motion equation of the scalar field is
\begin{equation}\label{57}
\begin{split}
&\frac{d}{dr}(\mathrm{\Delta}\frac{dR}{dr})+\\&[\frac{(k(r)+a^2)^2}{\mathrm{\Delta}}\omega^2-\frac{2a\eta}{\mathrm{\Delta}}m\omega+\frac{m^2a^2}{\mathrm{\Delta}}-\mu^2k(r)-\lambda_{lm}]R(r)=0.
\end{split}
\end{equation}

The angular part of the scalar field's equation of motion, equation \eqref{56}, is solved by spherical functions with eigenvalues $\lambda_{lm}$. Since later calculations of the energy-momentum tensor involve integration over the entire event horizon surface, and due to the normalization properties of spherical functions. Therefore, we do not solve specifically for the explicit form of the spherical functions here. Instead, our focus is on solving the radial equation.

To facilitate the solution of the radial equation, we introduce tortoise coordinates $r_\ast$ here.
\begin{equation}\label{58}
\frac{dr}{dr_\ast}=\frac{\mathrm{\Delta}}{k(r)+a^2}.
\end{equation}
Tortoise coordinates push the event horizon to infinity, so the radial coordinate covers the entire region of the spacetime from the event horizon outward. Substituting tortoise coordinates into the equation \eqref{57} and simplifies, we get
\begin{equation}\label{59}
\begin{split}
&\frac{\mathrm{\Delta}}{(k(r)+a^2)^2}\frac{d}{dr}(k(r)+a^2)\frac{dR}{dr_\ast}+\frac{d^2R}{dr_\ast^2}\\&+\left[\left(\omega-\frac{ma}{k(r)+a^2}\right)^2+\frac{\mathrm{\Delta}}{(k(r)+a^2)^2}2am\omega\right.\\
&\left.-\frac{\mathrm{\Delta}}{(k(r)+a^2)^2}(\mu^2k(r)+\lambda_{lm})\right]R(r)=0.
\end{split}
\end{equation}

We focus on the situation near the event horizon. In the vicinity, we can approximate and simplify the radial equation of the scalar field as follows
\begin{equation}\label{60}
\frac{d^2R}{dr_\ast^2}+\left(\omega-\frac{ma}{k(r)+a^2}\right)^2R(r)=0,
\end{equation}
substitute the angular velocity equation into the above formula to obtain
\begin{equation}\label{61}
\frac{d^2R}{dr_\ast^2}+\left(\omega-m\mathrm{\Omega}_h\right)^2R(r)=0,
\end{equation}
the solution of the above equation is
\begin{equation}\label{62}
R(r)=exp{\left[\pm i(\omega-m\mathrm{\Omega}_h)r_\ast\right]}.
\end{equation}
The positive and negative signs of the solution respectively represent the outgoing and incoming waves. Considering a scalar field entering a Kerr-like black hole, so the radial solution of the scalar field equation is
\begin{equation}\label{63}
R(r)=exp{\left[-i(\omega-m\mathrm{\Omega}_h)r_\ast\right]},
\end{equation}
substituting the above equation into the solution \eqref{55} of the scalar field equation, we obtain 
\begin{equation}\label{64}
\mathrm{\Psi}(t,r,\theta,\phi )=exp{\left[-i(\omega-m\mathrm{\Omega}_h)\right]}e^{-i\omega t}S_{lm}(\theta)e^{im\phi}.
\end{equation}

The above equation is the solution for the scalar field near the event horizon after a massive scalar field is incident on a Kerr-like black hole.

Here, we assume the mode of the scalar field is $(l,m)$. When the scalar field is incident on the Kerr-like black hole, part of its energy is absorbed by the black hole, while the rest is reflected back. In this context, our primary interest is in the absorbed portion of energy and angular momentum, exploring whether the black hole's event horizon can be disrupted after absorbing angular momentum and energy.

The energy-momentum tensor $T_{\mu\nu}$ for a scalar field with mass $\mu$ can be represented as
\begin{equation}\label{65}
T_{\mu\nu}=\partial_\mu\mathrm{\Psi}\partial_\nu\mathrm{\Psi}^\ast-\frac{1}{2}g_{\mu\nu}\left(\partial_\alpha\mathrm{\Psi}\partial^\alpha\mathrm{\Psi}^\ast+\mu^2\mathrm{\Psi}^\ast\mathrm{\Psi}\right).
\end{equation}
substituting the metric of a Kerr-like black hole into equation \eqref{65}, we obtain
\begin{equation}\label{66}
\frac{dE}{dt}=\int_{h}{T_t^r\sqrt{-g}d\theta d\phi=\omega(\omega-m\mathrm{\Omega}_h)\left(k(r)+a^2\right)},
\end{equation}
\begin{equation}\label{67}
\frac{dJ}{dt}=\int_{h}{T_\phi^r\sqrt{-g}d\theta d\phi=m(\omega-m\mathrm{\Omega}_h)\left(k(r)+a^2\right)}.
\end{equation}

Through the above two equations, it can be intuitively found that when $\omega>m\mathrm{\Omega}_h$, the angular momentum flux and energy flux through the event horizon are positive, indicating that the Kerr-like black hole absorbs energy and angular momentum from the scalar field. When $\omega<m\mathrm{\Omega}_h$, the energy flux and angular momentum flux through the event horizon are negative, signifying that the scalar field takes away the energy and angular momentum of the Kerr-like black hole, which is known as black hole superradiance\cite{Brito:2015oca}.

If we only consider a very short time interval $dt$, thus, we have
\begin{equation}\label{68}
dE=\omega(\omega-m\mathrm{\Omega}_h)\left(k(r)+a^2\right)\mathrm{dt},
\end{equation}
\begin{equation}\label{69}
dJ=m(\omega-m\mathrm{\Omega}_h)\left(k(r)+a^2\right)\mathrm{dt}.
\end{equation}

After calculating the changes in angular momentum and energy absorbed by the Kerr-like black hole from the scalar field, we can analyze whether the event horizon of the Kerr-like black hole, in extremal and near-extremal cases, can be disrupted by a scalar field with large angular momentum.

\subsection{Inducing the Kerr-like Black Hole Overspin by Injecting a Classical Scalar Field}
In this section, by injecting a monochromatic classical scalar field with frequency $\omega$ and angular quantum number $m$ into a Kerr-like black hole, we verify whether a scalar field with large angular momentum can disrupt the event horizon of such a black hole. During the scattering process, our focus is on the process within a small time interval $dt$. To verify the weak cosmic censorship conjecture, the sign of $M^{\prime2}-J\prime$ in the composite system needs to be considered. If $M^{\prime2}-J\prime\geq0$, implying that the weak cosmic censorship conjecture is not violated; if $M^{\prime2}-J\prime<0$, implying a violation of the weak cosmic censorship conjecture.

After the composite system absorbs the energy and angular momentum of the incident scalar field, the state of the composite system is
\begin{equation}\label{70}
M\prime^2-J\prime=(M^2-J)+2M\delta E+\delta E^2-\delta J ,
\end{equation}
considering the case of first-order perturbation, we have
\begin{equation}\label{71}
M\prime^2-J\prime=(M^2-J)+2M\delta E-\delta J .
\end{equation}
Substituting equations \eqref{68} and \eqref{69} into the above equation, we get
\begin{equation}\label{72}
M\prime^2-J\prime=(M^2-J)+2Mm^2(\frac{\omega}{m}-\frac{1}{2M})(\frac{\omega}{m}-{\Omega}_h)(k(r)+a^2)dt,
\end{equation}

When the initial state of the Kerr-like black hole is extremal, there is $J=M^2$, thus the above equation can be rewritten as
\begin{equation}\label{73}
M\prime^2-J\prime=2Mm^2(\frac{\omega}{m}-\frac{1}{2M})(\frac{\omega}{m}-{\Omega}_h)(k(r)+a^2)dt,
\end{equation}
the angular velocity $\mathrm{\Omega}_h$ of the Kerr-like black hole in the extremal case can be simplified to
\begin{widetext}
\begin{equation}\label{74}
\Omega _{h}=\frac{1}{2M+\sigma \left [ ln(1+\frac{M}{R_{s}})+\frac{R_{s}}{M+R_{s}}-1   \right ] +2\sqrt{\sigma M \left[ ln(1+\frac{M}{R_{s}})+\frac{R_{s}}{M+R_{s}}-1 \right ]}}.
\end{equation}
\end{widetext}

We inject the following scalar field modes into an extremal Kerr-like black hole
\begin{equation}\label{75}
\frac{\omega}{m}=\frac{1}{2}(\frac{1}{2M}+\mathrm{\Omega}_h).
\end{equation}
By substituting the above equation into the equation \eqref{74}, we obtain
\begin{equation}\label{76}
M\prime^2-J\prime=-\frac{1}{2}Mm^2(\mathrm{\Omega}_h-\frac{1}{2M})^2(k(r)+a^2)dt,
\end{equation}
According to equation \eqref{74}, the extreme Kerr-like black hole immersed in a dark matter halo has
\begin{equation}\label{77}
M\prime^2-J\prime\le 0.
\end{equation}

Clearly, when $\sigma \ne 0$, that is, $M\prime^2-J\prime < 0$, we find that in the first-order case, the event horizon of an extremal Kerr-like black hole can be destroyed. When $\sigma =0$ holds, i.e., $M\prime^2-J\prime = 0$, the black hole degenerates into a Kerr black hole. At this time, the event horizon of the extremal Kerr black hole cannot be disrupted. 

When the initial state of a Kerr-like black hole is in a near-extremal condition, i.e., $J\neq M^2$, we get
\begin{equation}\label{78}
M\prime^2-J\prime=(M^2-J)+2Mm^2(\frac{\omega}{m}-\frac{1}{2M})(\frac{\omega}{m}-{\Omega}_h)(k(r)+a^2)dt,
\end{equation}
the mode of the scalar field injecting in this near-extremal case is
\begin{equation}\label{79}
\frac{\omega}{m}=\frac{1}{2}(\mathrm{\Omega}_h+\frac{1}{2M}).
\end{equation}
Therefore, equation \eqref{78} can be expressed in the following form
\begin{equation}\label{80}
M\prime^2-J\prime=(M^2-J)-\frac{1}{8M}m^2\Omega _{h}^{2} \left ( \frac{1}{\Omega _{h}}-2M  \right ) ^2(k(r)+a^2)dt.
\end{equation}

Define a dimensionless small quantity $\varepsilon$, namely
\begin{equation}\label{81}
\frac{a^2}{M^2}=1-\varepsilon^2,
\end{equation}
shows that when the dimensionless small quantity $\varepsilon$ approaches zero, it describes a near-extremal situation; when the dimensionless small quantity $\varepsilon$ equals zero, it describes an extremal situation.

Calculated using the Taylor expansion
\begin{equation}\label{82}
M^2-J=\frac{M^2 \varepsilon ^2}{2} +O(\varepsilon ^4),
\end{equation}
\begin{widetext}
\begin{equation}\label{83}
\frac{1}{\Omega _{h}}-2M=k_{1}\sigma +2\sqrt{h_{1}}M+k_{1}\varepsilon +\left ( \frac{k_{1}\sigma }{2}+2M+\sqrt{h_{1}}M  \right )\varepsilon ^{2}+\frac{k_{1}}{2}\varepsilon ^{3} +O(\varepsilon ^{4}).
\end{equation}
\end{widetext}
Substituting the above equation into equation \eqref{80}, we obtain that in the near-extremal case, after a massive scalar field is absorbed by a Kerr-like black hole, the state of the resulting composite system becomes as follows
\begin{widetext}
\begin{equation}\label{84}
M\prime^2-J\prime=\left [ \frac{M^2 \varepsilon ^2}{2} +O(\varepsilon ^4) \right ] -\frac{1}{8M}m^2\Omega _{h}^2\left [ k_{1}\sigma +2\sqrt{h_{1}}M+k_{1}\varepsilon +\left ( \frac{k_{1}\sigma }{2}+2M+\sqrt{h_{1}}M  \right )\varepsilon ^{2}+\frac{k_{1}}{2}\varepsilon ^{3} +O(\varepsilon ^{4}) \right ]^2(k(r)+a^2)dt.
\end{equation}
\end{widetext}
Where
\begin{equation}\label{85}
k_{1}=ln\left ( 1+\frac{M+\varepsilon }{R_{s}}  \right )+\frac{R_{s}}{M+\varepsilon +R_{s}}-1,
\end{equation}
\begin{equation}\label{86}
h_{1}=\varepsilon ^2+\frac{\sigma (1+\varepsilon )\left [ ln\left ( 1+\frac{M+\varepsilon }{R_{s}}  \right )+\frac{R_{s}}{M+\varepsilon +R_{s}}-1  \right ] }{M}.
\end{equation}

Since a very short time interval $dt$ is considered, both $\varepsilon$ and $dt$ are first-order small quantities. As seen in the above equation, the first part is at most a second-order small quantity, while the largest part of the latter is a first-order small quantity. In this case, the overall sign presented by the equation is
\begin{equation}\label{87}
M\prime^2-J\prime<0.
\end{equation}
In other words, in near-extremal condition, the scalar field can disrupt the event horizon of a Kerr-like black hole. When there is no cold dark matter around the black hole, at this point, the leading part is at most a second-order small quantity, while the trailing part is at most a third-order small quantity. So, $M\prime^2-J\prime>0$, and the Kerr-like black hole degenerates into a Kerr black hole, meaning the event horizon of the near-extremal Kerr black hole cannot be disrupted by a scalar field. This is consistent with our previous research conclusions\cite{Meng:2023vkg}.

\section{Summary and discussion}
The veracity of the weak cosmic censorship conjecture is difficult to verify, people can only provide evidence for or against the conjecture in specific cases, and no general scheme to refute the conjecture has been proposed. Black holes existing in the universe are unlikely to be vacuum but filled with matter. Dark matter is a hot topic of research in this regard, so studying the Weak Cosmic Censorship Conjecture in the scenario of a rotating black hole enveloped by a dark matter halo will have more realistic physical significance.

In this paper, we obtained the first-order analytical solution of the black hole event horizon through approximate solving. To evaluate the reasonableness of the first-order analytical solution, we will provide the error range in the appendix. We conducted a study by injecting a test particle and a scalar field into the black hole. The results indicate that when using the test particle to verify the weak cosmic censorship conjecture, the test particle can disrupt the event horizon of a Kerr-like black hole in both extremal and near-extremal cases. For the scattering of the scalar field, our results show that the event horizon of a Kerr-like black hole can be disrupted in both extremal and near-extremal cases. Of course, when cold dark matter is not considered, the conclusion is consistent with previous research.

In summary, our results indicate that the presence of dark matter does lead to the over-rotation of rotating black holes. Of course, investigating the Weak Cosmic Censorship Conjecture in a more realistic black hole not only allows us to explore whether a rotating black hole can be over-spun but also provides a deeper understanding of dark matter. Moreover, for a black hole existing in the real universe, it serves as a natural laboratory because we can use particles in the accretion disk as test particles to investigate the Weak Cosmic Censorship Conjecture, thus gaining a deeper understanding of the conjecture.

\section*{Acknowledgements}
We acknowledge the anonymous referee for a constructive report that has significantly improved this paper. This work was supported by Guizhou Provincial Basic Research Program (Natural Science) (Grant No. QianKeHeJiChu-[2024]Young166), the Special Natural Science Fund of Guizhou University (Grant No.X2022133), the National Natural Science Foundation of China (Grant No. 12365008) and the Guizhou Provincial Basic Research Program (Natural Science) (Grant No. QianKeHeJiChu-ZK[2024]YiBan027).

\section*{Appendix}
\subsection{Error assessment}
For the approximate iterative method used in the second section of the article, we assess the reasonableness of using the first-order correction here. To evaluate the reasonableness, we introduce error $E=\frac{\left |  r_{+}^{num}-r_{+}^{app}\right | }{r_{+}^{num}} $ for assessment. $ r_{+}^{num}$ is the event horizon value calculated numerically, and $r_{+}^{app}$ is the event horizon value of the approximate analytical solution.

As shown in Tables \ref{tab:b} and \ref{tab:c}, since the effect of the dark matter halo is very small, we have quantitatively calculated the corresponding errors here. $\sigma =0.1$, $R_{s}=1.5$ is shown in Table \ref{tab:b}. $\sigma =0.2$, $R_{s}=1.5$ is shown in Table \ref{tab:c}. It is evident from Table \ref{tab:b} that the error of the first-order corrected analytical solution is of the order of $10^{-3}\sim 10^{-4}$, the error of the second-order corrected analytical solution is of the order of $10^{-4}\sim 10^{-5}$, and the error of the third-order corrected analytical solution is of the order of $10^{-5}$ or even smaller. For Table \ref{tab:c}, the error of the first-order corrected analytical solution is of the order of $10^{-2}\sim 10^{-3}$, the error of the second-order corrected analytical solution is of the order of $10^{-3}\sim 10^{-5}$, and the error of the third-order corrected analytical solution is of the order of $10^{-4}$ or even smaller. Therefore, using the first-order corrected analytical solution to explore whether the weak cosmic censorship conjecture can be violated is sufficient in this paper, as its accuracy can be above the order of $10^{-2}$.

\begin{table*}[ht]
    \centering
    \renewcommand{\arraystretch}{1.5} % 调整行间距，数值越大行间距越大
    \caption{By evaluating the error between the event horizon calculated through numerical methods and the event horizon calculated using iterative approximate analytical solutions, in the case of $\sigma =0.1$, $k=0.1$.}
    \begin{tabular}{p{1.0cm}p{1.7cm}p{1.7cm}p{1.8cm}p{1.7cm}p{1.8cm}p{1.7cm}p{1.8cm}p{1.7cm}p{1.7cm}}
        \hline\hline % 创建表格顶部的双线
       & \multicolumn{1}{l}{Numerical Solution} & \multicolumn{2}{l}{Zeroth-Order Correction} & \multicolumn{2}{l}{First-Order Correction} & \multicolumn{2}{l}{Second-Order Correction} & \multicolumn{2}{l}{Third-Order Correction} \\
        \hline
        \textbf{a} &  $r_{+}^{num}$ &  $r_{0+}$ & \textbf{Error} &  $r_{1+}$ & \textbf{Erro} & $r_{2+}$ & \textbf{Error} &  $r_{3+}$ & \textbf{Error} \\
        \hline
        0.1 & 2.02302 & 1.99499 & 0.0138555 & 2.02219 & 0.0004103 & 2.02299 & 0.0000148 & 2.02302 &0.0000000 \\
        0.3 & 1.98188 & 1.95394 & 0.0140977 & 1.98104 & 0.0004238 & 1.98185 & 0.0000151 & 1.98188 &0.0000000 \\
        0.5 & 1.89384 & 1.86603 & 0.0146845 & 1.89296 & 0.0004647 & 1.89381 & 0.0000158 & 1.89384 &0.0000000 \\
        0.7 & 1.74207 & 1.71414 & 0.0160327 & 1.74109 & 0.0005625 & 1.74203 & 0.0000230 & 1.74206 &0.0000057 \\
        0.9 & 1.46637 & 1.43589 & 0.0207860 & 1.46497 & 0.0009547 & 1.46630 & 0.0000477 & 1.46637 &0.0000000 \\
        0.99 & 1.19190 & 1.14107 & 0.0426462 & 1.18749 & 0.0037000 & 1.19151 & 0.0003272 & 1.19187 &0.0000252 \\
        0.999 & 1.13000 & 1.04471 & 0.0754779 & 1.11970 & 0.0091150 & 1.12874 & 0.0011150 & 1.12984 &0.0001416 \\
        \hline\hline % 创建表格底部的双线
    \end{tabular}
    \label{tab:b}
\end{table*}

\begin{table*}[ht]
    \centering
    \renewcommand{\arraystretch}{1.5} % 调整行间距，数值越大行间距越大
    \caption{By evaluating the error between the event horizon calculated through numerical methods and the event horizon calculated using iterative approximate analytical solutions, in the case of $\sigma =0.2$, $k=0.2$.}
    \begin{tabular}{p{1.0cm}p{1.7cm}p{1.7cm}p{1.8cm}p{1.7cm}p{1.8cm}p{1.7cm}p{1.8cm}p{1.7cm}p{1.7cm}}
        \hline\hline % 创建表格顶部的双线
       & \multicolumn{1}{l}{Numerical Solution} & \multicolumn{2}{l}{Zeroth-Order Correction} & \multicolumn{2}{l}{First-Order Correction} & \multicolumn{2}{l}{Second-Order Correction} & \multicolumn{2}{l}{Third-Order Correction} \\
        \hline
        \textbf{a} &  $r_{+}^{num}$ &  $r_{0+}$ & \textbf{Error} &  $r_{1+}$ & \textbf{Erro} & $r_{2+}$ & \textbf{Error} &  $r_{3+}$ & \textbf{Error} \\
        \hline
        0.1 & 2.05200& 1.99499 & 0.0277827 & 2.04869 & 0.0016131 & 2.05180 & 0.0000975 & 2.05198 & 0.0000096  \\
        0.3 & 2.01077& 1.95394 & 0.0282628 & 2.00740 & 0.0016760 & 2.01056 & 0.0001044 & 2.01075 & 0.0000100  \\
        0.5 & 1.92261& 1.86603 & 0.0294287 & 1.91911 & 0.0018204 & 1.92239 & 0.0001144 & 1.92259 & 0.0000104  \\
        0.7 & 1.77095& 1.71414 & 0.0320788 & 1.76709 & 0.0021796 & 1.77069 & 0.0001468 & 1.77093 & 0.0000113  \\
        0.9 & 1.49774& 1.43589 & 0.0412956 & 1.49234 & 0.0036054 & 1.49726 & 0.0003205 & 1.49770 & 0.0000267  \\
        0.99 & 1.23874& 1.14107 & 0.0788462 & 1.22451 & 0.0114875 & 1.23663 & 0.0017033 & 1.23842 & 0.0002583  \\
        0.999 & 1.18879& 1.04471 & 0.1211989 & 1.16327 & 0.0214672 & 1.18424 & 0.0038274 & 1.18798 & 0.0006814  \\
        \hline\hline % 创建表格底部的双线
    \end{tabular}
    \label{tab:c}
\end{table*}

%\nocite{*}
%\bibliographystyle{unsrt}
%\bibliography{ref}

\end{document}